\documentclass[12pt,a4paper]{article}
\usepackage[utf8]{inputenc}
\usepackage{geometry}
\usepackage{xurl}
\usepackage{graphicx}
\usepackage{setspace}
\usepackage{cite}
\usepackage[table]{xcolor}
\usepackage[most]{tcolorbox}
\usepackage{tabularx}
\usepackage{booktabs}
\usepackage{array}
\usepackage{ragged2e}
\usepackage{tikz}
\usepackage{placeins}
\usepackage{hyperref}
\usetikzlibrary{arrows.meta, positioning, shapes.geometric}

\newcolumntype{L}[1]{>{\RaggedRight\arraybackslash}p{#1}}
\newcolumntype{Y}{>{\RaggedRight\arraybackslash}X}

\title{\textbf{Beyond Critical Minerals Targets: Digital Rock Physics as Infrastructure for Secure and Circular Supply Chains}}
\author{%
Hannah P. Menke$^{1,2}$, Alessio Scanziani$^{3}$, Maja Rücker$^{4,5,6}$\\[0.7em]
\begin{tabular}{c}
\small $^{1}$Institute of GeoEnergy Engineering, Heriot-Watt University, Edinburgh, UK\\
\small $^{2}$Subsurface Energy Transition and Innovation Centre,\\
\small Heriot-Watt University, Edinburgh, UK\\
\small $^{3}$International Energy Agency, Paris, France\\
\small $^{4}$Eindhoven University of Technology, Mechanical Engineering,\\
\small 5600 MB, Eindhoven, The Netherlands\\
\small $^{5}$Eindhoven Institute for Renewable Energy Systems,\\
\small 5600 MB, Eindhoven, The Netherlands\\
\small $^{6}$Max Planck Institute for Polymer Research, 55128, Mainz, Germany
\end{tabular}}
\date{\today}

\begin{document}

\maketitle

\begin{abstract}
The United Kingdom and Europe are moving rapidly from critical-minerals target-setting to implementation. The EU Critical Raw Materials Act and the UK's Vision 2035 create ambitious benchmarks for domestic extraction, processing, recycling, circularity, and supply-chain resilience, but many prospective regional resources remain complex, underexplored, historically worked, or economically marginal. This paper argues that implementation will depend not only on permitting reform and project designation, but also on pre-competitive measurement, modelling, and data infrastructure capable of determining which ores, brines, waste streams, and recycling feedstocks can be processed viably and with lower environmental impact. Digital Rock Physics (DRP) should therefore be understood as enabling infrastructure for resource policy rather than as a specialist laboratory method alone. By combining three-dimensional imaging, correlative chemistry, AI-enabled image analysis, and pore-scale modelling, DRP can connect mineral texture and reactive pathways to decisions about ore characterisation, liberation prediction, leaching, Direct Lithium Extraction, mine-waste valorisation, and battery recycling. The paper sets out a UK-European policy agenda built around translational demonstrators, cross-disciplinary training, a Digital Ore Passport standard, a federated Digital Ore Database, and integrated geo-reactive endstations. Treated as shared implementation infrastructure, DRP could help turn critical-minerals strategies into practical routes for supply security, resource efficiency, circularity, and more environmentally responsible development.
\end{abstract}


\section{From Critical-Minerals Targets to Implementation}
The global transition to clean energy technologies has triggered rapidly rising demand for critical minerals such as copper, lithium, nickel, cobalt, rare earth elements (REEs), and graphite. Demand for other base metals such as aluminium and zinc, and strategic minor minerals such as titanium, tungsten, germanium, gallium, and antimony, is also rising because of expanding high-tech, aerospace, and defence sectors. Recent review literature places this challenge in a wider geopolitical and technological context, emphasizing both resource security and processing capability as core bottlenecks \cite{reich_critical_2025}. Global lithium demand, for example, surged by nearly 30\% in 2024, even as mining exploration activity plateaued and investment growth slowed markedly. Electrification of end uses is also driving increasing demand for copper; based on the current mine-project pipeline, a supply gap of up to 30\% could emerge by 2035, partly due to declining ore grades \cite{iea_outlook_2025}. Consequently, achieving higher recovery and better resource efficiency from existing deposits is increasingly important, particularly in regions where the central challenge is not the absolute absence of mineralisation but the difficulty of turning complex occurrences into robust projects.

Compounding this issue is the hyper-concentration of the market. The IEA reports that the average market share of the top three refining nations for key energy minerals rose to 86\% in 2024 \cite{iea_outlook_2025}. In response, the EU Critical Raw Materials Act (CRMA) established 2030 benchmarks requiring 10\% extraction, 40\% processing, and 25\% recycling to occur domestically \cite{eu_crma_2024}. Concurrently, the UK's strategy focuses on resilient supply chains, domestic capability, and leveraging world-leading academic strengths \cite{uk_strategy_2022}. Recent European compilation work identifies more than 800 medium- to very-large hard-rock CRM deposits across 33 countries, but also stresses that continental-scale resource assessment remains uneven and that many published figures should be treated as rough lower-end estimates rather than directly comparable reserve statements \cite{albert_crm_europe_2025}. In the UK, BGS has identified multiple prospective CRM regions while stressing that they remain underexplored and require much more systematic research before commercial potential can be judged \cite{bgs_uk_crm_2022}. The latest UK Vision 2035 therefore sets an ambition for at least 10\% of aggregate annual demand to be met through domestic production by 2035, underscoring both the scale of present dependence and the need to improve recovery wherever viable resources can be developed \cite{uk_vision_2035_2026}.

The resulting policy problem is no longer only whether governments can identify critical-mineral projects or set numerical supply targets. It is whether public policy can create the practical implementation layer needed to make complex resources, waste streams, and recycling pathways investable. Project designation, permitting reform, and strategic lists do not by themselves answer whether a low-grade ore can be liberated economically, whether a historic tailings deposit contains recoverable critical minerals, whether a brine or sorbent will retain selectivity under realistic operating conditions, or whether battery black mass can be routed to the most efficient recovery process. Those questions require shared measurement protocols, validated models, benchmark datasets, and institutions that allow evidence from laboratories, geological surveys, major facilities, and industrial pilots to be compared and reused. This is the policy gap addressed here.

The United States is also investing heavily in advanced mining innovation through a layered translation ecosystem: regional consortia for resource and feedstock development, long-lived hub-style programmes for early-stage research, cross-laboratory facilities such as METALLIC for modelling, data, and process integration, and Mine of the Future proving grounds and national-laboratory calls that bridge bench-scale work to field demonstration \cite{doe_regional_consortia_2025, doe_cmi_2025, doe_metallic_2024, doe_motf_launch_2025}. The IEA has also established a Technology Collaboration Programme on Critical Minerals and Materials Recovery to coordinate technical expertise and national R\&D efforts around secondary and unconventional mineral resources \cite{iea_cmmr_tcp_2026}. This pattern of investment and global interest in new recovery technologies is closely aligned with the requirements of DRP, whose value depends on the integration of multimodal imaging, correlative chemistry, modelling, data infrastructure, and process validation within a single translational workflow. The UK and Europe are well placed to draw on strong existing DRP communities in academia and major facilities, to extend transatlantic collaboration, and to define a distinctive role in the critical-minerals technology stack. Europe has moved decisively on project designation and permitting through the CRMA, and the UK now signals additional support through Vision 2035, but a comparably visible experimental-and-modelling translation stack remains less developed \cite{eu_strategic_projects_2025, uk_vision_2035_2026}. For the UK and Europe, the issue is not simple replication of the US model, but the development of a credible programme built on existing strengths in synchrotron science, applied mineralogy, and geoscience infrastructure. Many prospective UK and European deposits are geologically heterogeneous, historically worked, or unlikely to compete on grade alone with the largest global suppliers. Moreover, the UK and Europe are densely populated, requiring particular attention to environmental impact, operational safety, and efficiency. In such settings, sustainability must be integrated into project design from the outset, both to reduce environmental burden and to maintain the social licence on which new supply depends. Despite the numerous challenges, geological data indicate that Europe and the UK still has significant untapped opportunities for critical minerals extraction. For example, for lithium, graphite and nickel the regions combined account for 2\%, 3\% and 4\% respectively of global reserves, while current production is less than 1\% for lithium and graphite, and less than 2\% for nickel \cite{iea_outlook_2025}. Under those conditions, methods that improve orebody discrimination, liberation prediction, leach efficiency, recycling yields, or by-product determination by a few percentage points have strategic value \cite{albert_crm_europe_2025, bgs_uk_crm_2022, calvo_copper_2016}.

\section{Digital Rock Physics as Enabling Infrastructure}
Digital Rock Physics (DRP) is best understood here not as a single technique, but as a shared evidence system for linking three-dimensional material structure to recovery, processing, and environmental performance. Its technical base includes X-ray micro-computed tomography ($\mu$CT), nano-CT, PET, MRI, NMR, FIB-SEM, helium ion microscopy, correlative chemistry, image processing, AI-driven segmentation, and multi-scale modelling \cite{drp_blunt_2013}. For resource policy, however, the important point is not the instrument list. It is that these tools can generate comparable, reusable evidence about texture, connectivity, mineral chemistry, reactive pathways, and process response in materials that conventional bulk assays or two-dimensional sections cannot fully resolve.

Historically, this toolkit was developed to optimise hydrocarbon extraction and later CO$_2$ sequestration. However, the same methods are well suited to a broader class of mineral-systems problems in which three-dimensional texture, connectivity, and evolving fluid-mineral interactions control recovery. The scientific study of porous materials, including through DRP, is closely intertwined with multiple energy applications, because porous materials can mediate the transfer of mass, charge, heat, radiation, and pressure across length, time, and energy scales \cite{farber_science_2025}. As the energy transition accelerates, this existing infrastructure, including synchrotrons, lab-scale scanners, and supercomputing clusters, is well positioned for reorientation toward hard-rock mining and mineralogy.

Recent reviews of integrated geometallurgy and applied mineralogy across the mine life cycle establish the importance of mineralogical heterogeneity, texture, and sustainability \cite{lishchuk_integrated_2020, becker_applied_2023}. DRP extends those concerns to specific technical decisions by combining analytical tools with imaging, AI-assisted analysis, multi-scale flow simulation, and data infrastructure within a single workflow.

DRP also occupies a different role from the automated mineralogy tools already deployed at industrial scale. SEM-based energy-dispersive spectroscopy platforms such as QEMSCAN, MLA, and TIMA are the current standard for mineral identification, liberation analysis, and association mapping in operating mines and processing plants, providing rapid, high-throughput chemical classification from polished sections. Because they are two-dimensional and destructive by sample preparation, they cannot capture 3D grain topology, pore-network connectivity, or dynamic behaviour during leaching, flotation, or reactive transport. DRP, particularly when it combines 3D X-ray CT with correlative chemical mapping and pore-scale flow simulation, supplies that structural and dynamic context: true liberation volumes rather than section-inferred estimates, connected flow pathways through particle beds or heap structures, and time-resolved changes in pore geometry and mineral chemistry during processing. In practice the two are typically used together, with automated mineralogy providing throughput and chemical classification, and DRP providing the 3D textural and dynamic evidence needed to explain why a given ore or material behaves as it does under processing conditions.

The UK and Europe already host much of the infrastructure needed for this shift, and that infrastructure is more distributed than a synchrotron-centred account might imply. In the UK, the EPSRC-supported NXCT facility links Manchester, Southampton, UCL, Warwick, and Diamond as a national lab-based X-ray CT network, while Manchester's Henry Moseley X-ray Imaging Facility alone hosts ten CT scanners, in situ rigs, and dedicated 3D analysis capability \cite{manchester_nxct_2026}. Across Europe, the EXCITE transnational-access programme connects a broader set of electron and X-ray imaging laboratories, and infrastructures such as ECCSEL extend lab-scale XCT and in situ experimental capability for geomaterials, reactive flow, and mechanical testing \cite{excite_access_2026, eccsel_strath_xct_2026}. What remains limited is coordination, shared data standards, and translation into mineral-processing workflows.

Nor is that regional base limited to structure-only imaging. Some centres already provide explicitly correlative or chemistry-aware workflows: UCL's Centre for Correlative X-ray Microscopy couples multi-scale XCT to wider materials-characterisation infrastructure, while CEITEC combines micro- and nano-CT with LIBS-based elemental mapping \cite{ucl_correlative_xray_2026, ceitec_xct_2026}. Major facilities extend that chemistry further. Diamond's I18 and I14 beamlines support micro- to nano-scale XRF/XANES/XRD mapping and combined experiments such as XRF tomography, XAS-CT, and XRD-CT, while DIAD enables quasi-simultaneous imaging and diffraction under user-supplied in situ conditions \cite{diamond_xrf_2026, diamond_diad_2026}. ESRF similarly spans chemically sensitive nano-imaging at ID16B, operando diffraction and imaging at ID15A, and X-ray absorption/emission spectroscopy at ID26 with gas handling, FT-IR, and gas-chromatography mass spectrometry \cite{esrf_id16b_2026, esrf_id15a_2026, esrf_id26_2026}. PSI's TOMCAT remains a leading dynamic imaging platform, while MAX IV already supports simultaneous in situ XAS-XRD at Balder and is explicitly developing TomoWISE and SpectroWISE as complementary structure-and-chemistry investments \cite{psi_tomcat_2025, maxiv_balder_2026, maxiv_beamlines_2024}.

A capability gap nevertheless remains. Across this distributed landscape, routine provision is still strongest either in structure-rich imaging or in chemistry-rich spectroscopy, but only rarely in synchronised measurement of 3D texture, mineral chemistry, reacting fluids, dissolved products, and mechanically evolving microstructure within a single experiment. ESRF's European Battery Hub addresses an analogous problem by coordinating multiple beamlines and methods around one materials system rather than relying on a single instrument \cite{esrf_eubat_2026}. Critical-minerals research requires comparable multimodal coordination, but with sample environments designed for heterogeneous ores, tailings, brines, and black-mass feedstocks rather than the more controlled materials used in batteries or model catalysts, and with control of flow, temperature, pressure, and mechanical loading where those conditions govern breakage, permeability evolution, or sorbent degradation.

The capability base therefore already exists; the main limitation is the weak integration of 3D structure, mineral chemistry, reactive transport, and reusable data standards within ore-system workflows.

\begin{figure}[htbp]
\centering
\includegraphics[width=\textwidth]{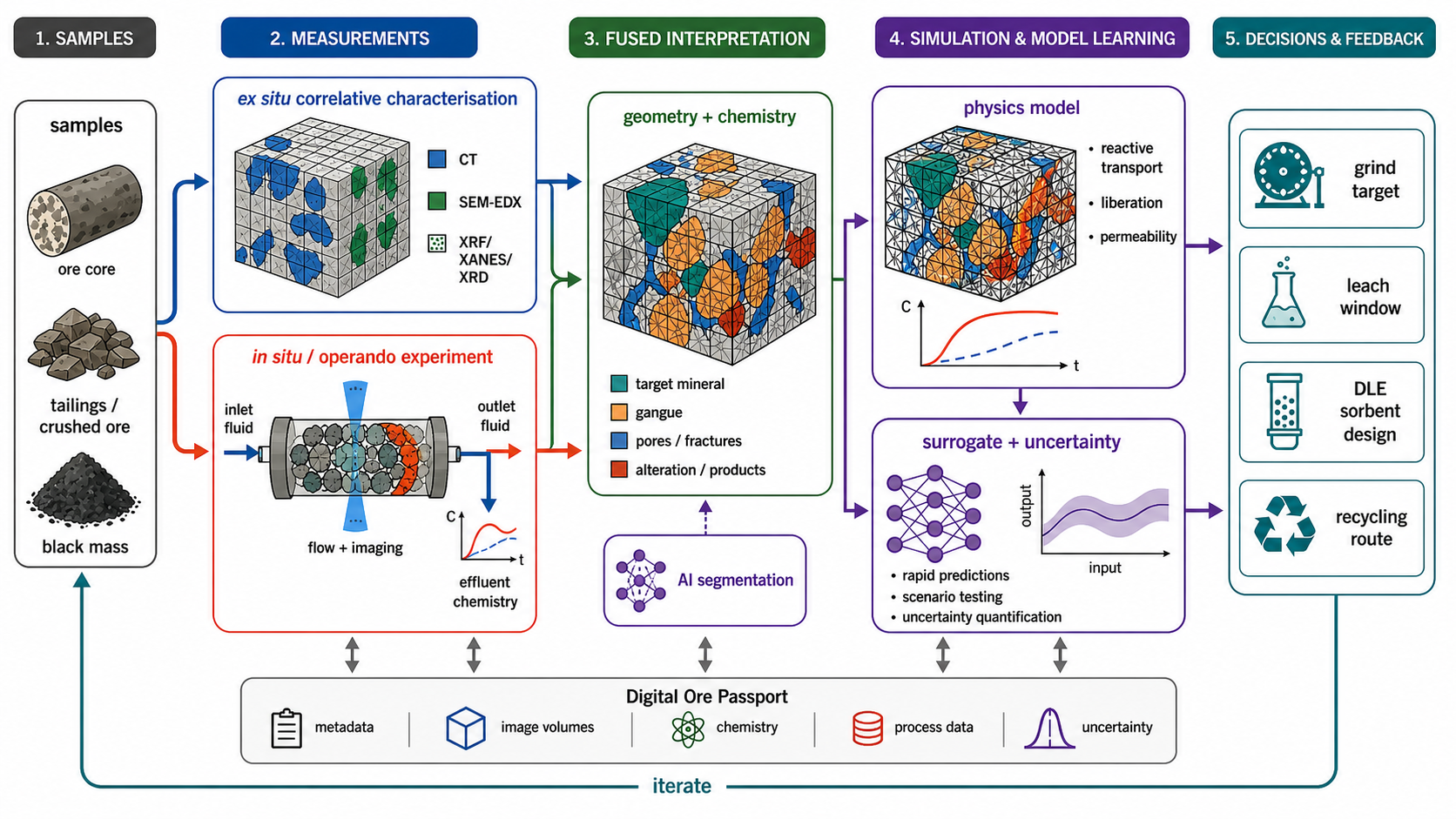}
\caption{Integrated Digital Rock Physics workflow for critical-mineral systems. Static correlative imaging resolves registered geometry and mineral chemistry, while in situ experiments track reactive flow, structural change, and effluent chemistry under process-relevant conditions. AI-assisted segmentation supports image interpretation, physics-based models translate those data into transport, liberation, and reaction predictions, and surrogate models with uncertainty estimates accelerate decision support for comminution, leaching, Direct Lithium Extraction, recycling, and tailings valorisation.}
\label{fig:drp_integrated_workflow}
\end{figure}

Figure~\ref{fig:drp_integrated_workflow} summarises the integrated DRP workflow proposed here, linking ex situ correlative characterisation, in situ reactive experiments, image interpretation, physics-based modelling, AI-enabled uncertainty analysis, and process decision-making.

\begin{figure}[htbp]
\centering
\resizebox{\textwidth}{!}{%
\begin{tikzpicture}[
  stage/.style={rectangle, rounded corners=4pt, draw=black!35, minimum width=3.4cm,
                minimum height=1.1cm, align=center, font=\bfseries\small, inner sep=5pt},
  contrib/.style={rectangle, rounded corners=4pt, draw=black!20,
                  minimum width=3.4cm, text width=3.2cm,
                  align=left, font=\footnotesize, inner sep=6pt},
  steparrow/.style={-{Stealth[length=7pt,width=5pt]}, line width=1.2pt, draw=black!45},
  downarrow/.style={-{Stealth[length=5pt,width=4pt]}, draw=black!30, dashed},
]

\node[stage, fill=blue!10]   (EXP)  at ( 0.00, 0) {I Exploration};
\node[stage, fill=cyan!10]   (CHAR) at ( 4.50, 0) {II Characterisation};
\node[stage, fill=orange!12] (PROC) at ( 9.00, 0) {III Processing \&\\[-2pt]Extraction};
\node[stage, fill=violet!8]  (ENGI) at (13.50, 0) {IV Engineered\\[-2pt]Materials};
\node[stage, fill=green!10]  (REC)  at (18.00, 0) {V Recycling \&\\[-2pt]Tailings};

\draw[steparrow] (EXP.east)  -- (CHAR.west);
\draw[steparrow] (CHAR.east) -- (PROC.west);
\draw[steparrow] (PROC.east) -- (ENGI.west);
\draw[steparrow] (ENGI.east) -- (REC.west);

\node[contrib, fill=blue!4, below=1.0cm of EXP] (EXP_C) {%
  \textbf{Methods:}\\
  $\mu$CT, nano-CT,\\
  AI segmentation\\[3pt]
  \textbf{Outputs:}\\
  3D ore-texture priors\\
  Prospectivity training data\\
  Rapid complex-ore screening\\[3pt]
  \textbf{Decision:}\\
  Drill-target ranking \&\\
  resource estimation
};

\node[contrib, fill=cyan!4, below=1.0cm of CHAR] (CHAR_C) {%
  \textbf{Methods:}\\
  $\mu$CT + SEM-EDX,\\
  FIB-SEM, XRF-CT\\[3pt]
  \textbf{Outputs:}\\
  3D liberation metrics\\
  Mineral associations\\
  Grain-size \& texture maps\\[3pt]
  \textbf{Decision:}\\
  Comminution target \&\\
  flowsheet design
};

\node[contrib, fill=orange!6, below=1.0cm of PROC] (PROC_C) {%
  \textbf{Methods:}\\
  PET, MRI, $\mu$CT,\\
  pore-scale simulation\\[3pt]
  \textbf{Outputs:}\\
  Leach flow-path maps\\
  Permeability evolution\\
  Breakage pathway models\\[3pt]
  \textbf{Decision:}\\
  Heap/leach optimisation \&\\
  reagent targeting
};

\node[contrib, fill=violet!4, below=1.0cm of ENGI] (ENGI_C) {%
  \textbf{Methods:}\\
  Nano-CT, NMR,\\
  in situ cycling CT\\[3pt]
  \textbf{Outputs:}\\
  Sorbent pore architecture\\
  Accessibility \& tortuosity\\
  Degradation \& pore blockage\\[3pt]
  \textbf{Decision:}\\
  MOF/membrane design \&\\
  DLE process conditions
};

\node[contrib, fill=green!4, below=1.0cm of REC] (REC_C) {%
  \textbf{Methods:}\\
  FIB-SEM, nano-CT,\\
  automated mineralogy\\[3pt]
  \textbf{Outputs:}\\
  Black-mass phase maps\\
  Tailings liberation texture\\
  Sensor calibration targets\\[3pt]
  \textbf{Decision:}\\
  Recycling route selection \&\\
  tailings flowsheet design
};

\draw[downarrow] (EXP.south)  -- (EXP_C.north);
\draw[downarrow] (CHAR.south) -- (CHAR_C.north);
\draw[downarrow] (PROC.south) -- (PROC_C.north);
\draw[downarrow] (ENGI.south) -- (ENGI_C.north);
\draw[downarrow] (REC.south)  -- (REC_C.north);

\end{tikzpicture}%
}
\caption{Digital Rock Physics contributions across the critical-minerals value chain. Top row: value chain stages linked by process arrows. Bottom row: the specific DRP imaging methods, structural and chemical outputs, and process or investment decisions those outputs inform at each stage. The ``Engineered Materials'' stage reflects the growing role of DRP in designing porous sorbents, MOFs, and ion-selective membranes for Direct Lithium Extraction and related low-footprint separation technologies.}
\label{fig:drp_value_chain}
\end{figure}
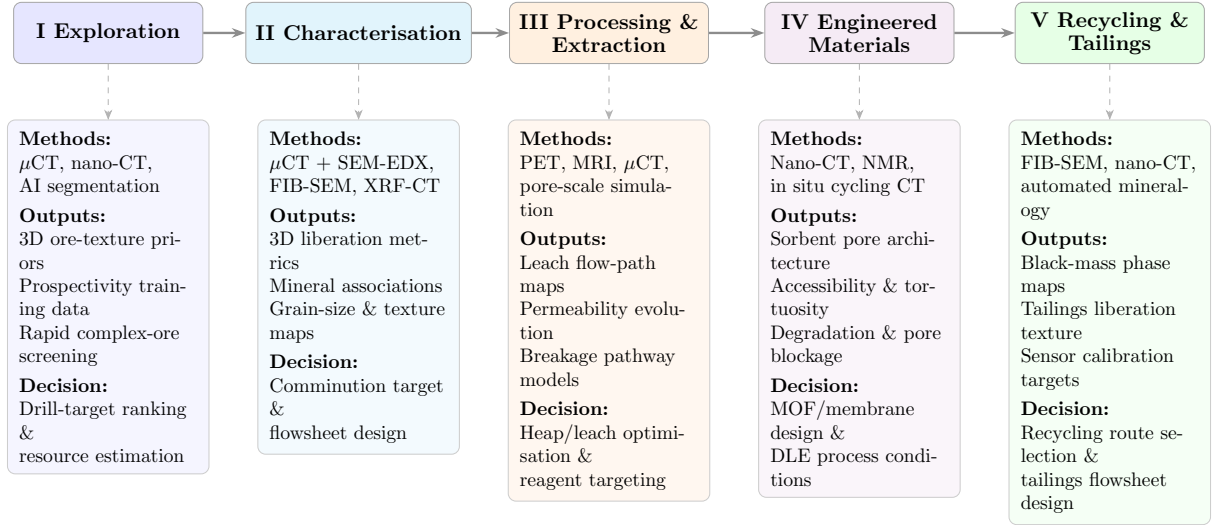

Figure~\ref{fig:drp_value_chain} shows where DRP can add decision-relevant information across the critical-minerals value chain, from exploration through recycling, together with the specific imaging methods and process decisions that DRP informs at each stage.

Modelled on the Digital Porous Media Portal, formerly the Digital Rocks Portal, a federated European ``Digital Ore Database'' sharing multi-modal tomographic data on critical-mineral deposits would broaden access to high-resolution structural data \cite{digital_porous_media_portal_2025, digitalrocks_portal_2023}.

Table~\ref{tab:mineral_categories} maps broad strategic categories of critical minerals to representative examples and indicates where DRP may have the greatest practical relevance in UK and European settings.

\begin{table}[htbp]
\centering
\footnotesize
\setlength{\tabcolsep}{6pt}
\renewcommand{\arraystretch}{1.25}
\rowcolors{2}{black!3}{white}
\begin{tabularx}{\textwidth}{L{0.18\textwidth}L{0.19\textwidth}Y Y}
\toprule
\textbf{Category} & \textbf{Example Minerals} & \textbf{Why DRP Helps} & \textbf{Why Relevant in UK/Europe} \\
\midrule
Battery materials & Lithium, cobalt, nickel, manganese, graphite & 3D texture mapping, liberation analysis, pore-scale transport, black-mass characterisation, and porous sorbent design for DLE & Cornwall's hard-rock and brine lithium resources, including the St Austell district and geothermal brine projects; Portugal's Barroso spodumene pegmatites; Finland's Keliber lithium project; Finnish nickel-cobalt sulphides (Terrafame, Sotkamo); and EU battery recycling streams \\
Magnet and high-tech materials & Rare earth elements, gallium, germanium & Correlative imaging for finely disseminated phases, automated mineralogy, and selective separation workflow design & Sweden's Per Geijer iron-apatite deposits (LKAB) and Norra Kärr alkaline complex; Norway's Fen carbonatite REE province; downstream gallium and germanium recovery from zinc smelter residues in Belgium and Germany \\
Technology and alloy metals & Tungsten, tin, antimony & Ore-texture characterisation, selective breakage studies, tailings re-evaluation, and sensor calibration for sorting & Cornwall's South Crofty tin mine and Drakelands tungsten deposit; Portugal's Panasqueira tungsten-tin mine; antimony-bearing tailings across Iberia and the UK's southwest mining districts \\
Secondary resources & Tailings, battery black mass, industrial residues & Multi-scale characterisation of heterogeneous feedstocks, recovery-pathway design, and calibration of sorting and recycling processes & EU battery gigafactory waste streams; Cornish and Iberian historical mining tailings; aluminium and copper smelter by-products in Germany, France, and Belgium, where urban and industrial recycling aligns with CRMA circularity targets \\
\bottomrule
\end{tabularx}
\caption{Representative mineral categories and examples illustrating where DRP can add value across UK and European resources.}
\label{tab:mineral_categories}
\end{table}

The most credible early applications are already visible. They include lithium-favourable pegmatite terrains in Iberia and elsewhere in Europe, south-west England's prospective lithium and established tin-tungsten province, Nordic and wider European CRM provinces captured on EuroGeoSurveys maps, and battery-recycling streams in which black mass is already being treated as a strategic secondary feedstock in EU projects \cite{egs_lithium_map_2020, bgs_uk_crm_2022, egs_crm_map_2016, batraw_2022}. These are plausible settings in which a DRP-led demonstrator programme could begin quickly and show practical value.

\section{Pillars I and II: Exploration and Advanced Characterisation}
DRP can reorient mineral exploration strategies by providing the three-dimensional ground-truth datasets that AI-enabled prospectivity mapping requires. Recent review literature identifies this as a rapidly developing frontier, with mineral-prospectivity mapping among its best-established applications \cite{yang_ai_exploration_2024, ai_mineral_prospectivity_2011}. It can also shorten the path from discovery to commercial evaluation by enabling faster orebody assessment using advanced sensing and analytical methods \cite{doe_arpae}.

Moving beyond traditional 2D thin sections, multi-modal tomographic approaches from macro-CT to nano-CT can quantify 3D mineral associations, grain sizes, and liberation characteristics. When paired with SEM-EDX-based automated mineralogy, these correlative workflows can locate finely disseminated or texturally complex critical minerals that might otherwise escape detection \cite{buyse_correlative_2023}. Emerging 3D liberation models demonstrate that 2D section-based estimates, which remain the current default in automated mineralogy, systematically misrepresent true liberation distributions for texturally complex ores, because a single polished section cannot capture the three-dimensional grain topology that controls whether a mineral particle is actually free, locked, or middling after breakage. In a recent direct comparison of SEM-based 2D image analysis with 3D X-ray microtomography, 2D liberation overestimated overall iron-oxide liberation by 9\% in one low-grade ore and by 22\% in another; in the more biased case, 2D values of 84--97\% contrasted with 3D maxima of only 75--79\% \cite{guntoro_liberation_2021, uliana_liberation_2025}. Together, these data-rich models support predictive geometallurgy by integrating sparse process data into spatial orebody models before full-scale mining, improving flowsheet design from the outset \cite{lishchuk_predictive_2019}. In operating comminution circuits, that predictive gain is already measurable: a geometallurgical throughput model built from production data at Tropicana performed 6.3\% better when hardness was represented compositionally rather than by average penetration rate, and adding comminution variables reduced throughput-prediction RMSE by a further 10.6\% \cite{both_geomet_2021}.

\begin{tcolorbox}[enhanced, colback=blue!3!white, colframe=blue!35!black, title=\textbf{Illustrative Case Study: Correlative Imaging for Complex Pegmatites}, fonttitle=\bfseries, arc=1.5mm, boxrule=0.8pt]
Buyse et al. combined X-ray micro-CT with SEM-EDX automated mineralogy to resolve internal mineral relationships in Sn-Nb-Ta pegmatites and build a mineral library for 3D interpretation \cite{buyse_correlative_2023}. Similar workflows can be adapted to lithium-bearing pegmatites and other texturally complex ores for which 2D sections provide only a partial guide to processing behaviour. In practice, this reduces reliance on 2D inference and provides a firmer basis for liberation testing and flowsheet design.
\end{tcolorbox}

\section{Pillars III and IV: Optimising Mineral Processing, Extraction, and Engineered Materials}
Critical mineral extraction is energy- and water-intensive at scale, and DRP offers several pathways for optimising these processes. Comminution (crushing and grinding), for example, is exceptionally energy-intensive: grinding alone accounts for approximately 3\% of global electricity generation and routinely exceeds 50\% of a single mine site's total power demand \cite{norgate_comminution_2010}. At the ore and process-test scale, improvements such as a roughly 14\% reduction in relative grindability index, from 0.93 to 0.80, to reach the same 80\% passing 74~$\mu$m grind size after inducing preferential intergranular breakage in magnetite ore, together with a 13.19\% increase in overall liberation, point to a plausible route for reducing comminution energy demand and improving downstream recovery if those effects persist under plant-scale operating conditions \cite{kyle_processing_2015, gao_hvep_2017}. DRP enables the microstructural characterisation of grain boundaries, fracture networks, and mineral associations that underpin these more selective strategies. When those image-derived descriptors are linked to ML models trained on process and assay data, the workflow can shift from retrospective diagnosis toward forward process optimisation, including adaptive grinding targets, breakage prediction, and faster flowsheet screening.

Similarly, in operations such as in-situ and heap leaching, used for example in the recovery of copper from low-grade ores, engineers can use dynamic imaging (e.g., PET or MRI) and pore-scale flow simulation to map leach-solution transport \cite{mri_heap_leach_2015}. That kind of process understanding can improve permeability management and dissolution efficiency while reducing unnecessary excavation. In this setting, surrogate models and physics-informed ML can be trained against time-resolved image volumes, effluent chemistry, and operating parameters to accelerate scenario testing and identify the variables most likely to improve contact efficiency or selective dissolution. DRP is also relevant to emerging direct extraction technologies. Engineered porous materials such as metal-organic frameworks (MOFs), ion sieves, and ion-selective membranes are increasingly being explored for lower-footprint separations, including lithium extraction from brines \cite{mofs_lithium_2026, mojid_dle_2024, baird_dle_2024}. In these systems, the key imaging target is often not the brine itself but the pore architecture, accessibility, degradation, and cycling behaviour of the sorbents or membranes that control selectivity and throughput; AI-assisted segmentation and reduced-order models can then be used to connect those structural changes to capacity fade, fouling, and operating windows.

In situ DRP is most straightforward in fixed-solid or slowly evolving flow-through systems, where fluid movement can be interpreted against a stable pore structure. Agitated particle systems require substantially faster imaging and time-resolved analysis because the geometry, contacts, and transport pathways evolve continuously during measurement.

Across these examples, DRP shifts process decisions that would otherwise rest on bulk assay and empirical trial onto a microstructural footing. In comminution, this means targeting preferential breakage along grain boundaries rather than minimizing particle size indiscriminately, thereby reducing energy input while improving liberation. In heap leaching, it means identifying poorly contacted and preferential-flow regions before they propagate through a full-scale heap, where rectification is costly or impossible: MRI measurements in a drip-irrigated laboratory leach column found that more than 99\% of the liquid remained within 2 mm of the ore surface, while pore space 2--3 mm from the solid was only 2.0--4.4\% liquid-filled, indicating strongly localized wetting rather than uniform solution distribution \cite{mri_heap_leach_2015}. In DLE, it means detecting the early-stage pore blockage, phase transformation, or mechanical degradation in sorbent materials before measurable capacity loss appears at pilot scale, providing diagnostic information early enough to alter operating conditions or sorbent design. Just as importantly, these capabilities support more defensible environmental decisions by revealing where energy, water, and reagents are being used inefficiently and where lower-impact processing routes may be feasible. Structural and chemical evidence at the micron to sub-micron scale can therefore resolve process ambiguities that bulk measurements leave open, while AI-enabled segmentation, surrogate modelling, and experimental design can carry those insights into operating choices while intervention is still practical.

Similarly, DRP insights can be extended to assess engineered materials derived from extracted products, such as porous battery electrodes, helping relate extraction quality directly to downstream product performance. In battery manufacturing, for example, the electrochemical performance of lithium-ion battery electrodes is strongly governed by their three-dimensional microstructure, including porosity, tortuosity, and phase connectivity. DRP techniques can therefore provide high-value insights for material design and performance optimisation while supporting the efficient use of critical minerals such as graphite, nickel, and cobalt.

\section{Pillar V: Tailings Valorisation and the Circular Economy}
Legacy mine tailings can contain unrecovered critical minerals (tellurium, antimony, and tungsten among them) at concentrations that may be economically significant in favourable site-specific settings; recovering them is central to both UK and EU circularity strategies. Recent reviews highlight both the strategic importance of reprocessing mine tailings and the opportunity to recover critical materials from them \cite{sarker_tailings_2022, araya_tailings_2020, nwaila_minewaste_2021}. 

To recycle these materials effectively, advanced sorting mechanisms need to be deployed and calibrated. Sensor-based sorting reviews in mineral processing point to preconcentration, waste reduction, and multi-sensor fusion as key opportunities for more selective treatment of heterogeneous feedstocks \cite{peukert_sorting_2022}. High-resolution 3D imaging can serve as a micro-scale calibration tool for these industrial sensors, while ML classifiers can use those image-derived labels to improve sensor fusion, particle discrimination, and route selection across heterogeneous waste streams. Multi-scale DRP methods can also be applied to unconventional secondary feedstocks such as coal refuse, fly ash, and acid mine drainage. In addition, techniques such as FIB-SEM and nano-CT can characterise e-waste and battery ``black mass'' at sub-micron scales to optimise both mechanical and chemical recycling pathways by mapping structural degradation and elemental composition \cite{battery_blackmass_2021}; in this setting, AI methods are well suited to phase identification, anomaly detection, and linking highly variable textures to downstream recovery performance.

The central technical challenge in tailings and secondary resources is heterogeneity across multiple scales. Unlike primary ores, which retain the mineralogical architecture of the original deposit, tailings have been crushed, ground, chemically modified by reagents, and weathered over decades. The result is a highly variable feedstock in which critical mineral phases may be finely disseminated, encapsulated in gangue, partially oxidised, or intermixed with processing reagent residues. Conventional bulk assay can confirm that value is present but cannot resolve whether it is recoverable under a given process, at what liberation size, or with what lixiviant chemistry. Multi-scale DRP, combining macro-CT for particle-level sorting with FIB-SEM for phase-association and alteration mapping, provides the textural evidence needed to design a reprocessing flowsheet for a specific tailings deposit rather than applying a generic process template. AI extends that workflow by clustering particle populations, ranking recovery routes, and learning from linked texture-process-outcome datasets that are too heterogeneous for manual interpretation alone. The economic consequence is clear: flowsheet overdesign for heterogeneous tailings wastes reagent and energy, while underdesign leaves value in the discard.

Because DRP reveals multiphysics and multiscale processes across mineral extraction, material design and manufacturing, and waste management, it could also complement life-cycle analysis (LCA). DRP may contribute to estimates of extraction efficiency, material efficiency, and energy efficiency, as well as the identification and quantification of potential side effects. Comparative standards in this domain would enable more predictive assessment of critical-mineral value chains before they are realised. By supporting considerations such as co-extraction of multiple minerals, integrated subsurface valorisation, for example through coupling with geothermal applications, and application-driven purification, DRP could contribute to the more efficient and sustainable use of mineral resources.

\begin{tcolorbox}[enhanced, colback=green!3!white, colframe=green!35!black, title=\textbf{Illustrative Case Study: Black-Mass Characterisation}, fonttitle=\bfseries, arc=1.5mm, boxrule=0.8pt]
Vanderbruggen et al. showed that automated mineralogy can characterise the composition and texture of spent lithium-ion batteries in ways that are directly relevant to recycling process design \cite{battery_blackmass_2021}. For a DRP-led critical-minerals agenda, this matters because it moves ``urban mining'' from bulk assay toward texture-aware process optimisation, where liberation, coating residues, and phase associations all influence recovery. The practical outcome is better discrimination between recycling routes and a clearer basis for process tuning.
\end{tcolorbox}

\section{Correlative Chemistry, AI, and Broader Relevance}
A major development in DRP is the integration of chemistry with structure within the same three-dimensional volume. Multi-modal workflows that combine CT, automated mineralogy, elemental mapping, and diffraction make it possible to follow mineral texture, chemical state, and reactive change together rather than as separate observations \cite{buyse_correlative_2023, koch_geomet_2019, gu_automated_mineralogy_2014}. This approach extends from geological samples to engineered porous materials such as MOFs, ion-selective membranes, and battery electrodes, where pore architecture, accessibility, and degradation control performance in DLE and related separation technologies \cite{farber_science_2025}.

AI and numerical modelling are equally central to this development. Open image-based solvers such as GeoChemFoam already show how reactive transport can be modelled directly in tomographic geometries and then used to construct more faithful upscaled descriptions \cite{maes_geochemfoam_2021}. Image-based surrogates, physics-informed ML, and benchmarked workflows offer a route to faster prediction, provided that physical consistency, uncertainty, and auditability are treated as part of the method rather than deferred to later validation \cite{graczyk_dl_porosity_2020, karniadakis_piml_2021, menke_geoenergy_ml_2026}. This is particularly important for the upscaling problem, where microscopic image volumes must ultimately inform decisions at the scale of cores, process units, or mine sites \cite{ml_upscaling_drp_2024, menke_upscaling_2021}.

These capabilities also have relevance beyond ore processing narrowly defined. In mafic and ultramafic systems, the same reactive imaging platforms that can interrogate leaching, permeability evolution, and sorbent degradation are also directly relevant to carbon mineralisation and other subsurface geoenergy applications \cite{matter_carbfix_2009, menke_dynamic_reaction_2015, menke_hydrogen_2024}. Fast imaging and pore-scale analysis likewise improve understanding of wettability and multiphase flow, with transferability across flotation, leaching, and geological storage problems \cite{bartels_fast_xray_2017, garfi_wetting_2022, berg_interface_darcy_2026}. This broader applicability strengthens the case for shared infrastructure by linking critical-minerals research to adjacent experimental and modelling communities.

\section{Policy Challenges and Recommendations}
To realise the potential of DRP in critical minerals, several hurdles still need to be addressed. A major limitation is the lack of published, head-to-head benchmarks comparing DRP-informed decisions with conventional practice on real ore systems. In oil and gas, the case for DRP was built progressively through such comparisons, showing, for example, that pore-scale simulation could match core-flood measurements and that image-derived permeability predictions could hold at field scale. An equivalent evidence base for critical minerals has not yet appeared in the open literature at the scale needed to inform investment decisions. The quantitative projections offered here therefore rest on physical reasoning and analogy with oil and gas applications rather than on independently validated industry benchmarks. Standardisation is also needed, particularly in imaging protocols and machine-learning models across the European research and industrial community. Recent work on ML in sustainable geoenergy makes the same broader point: benchmark design, uncertainty reporting, and auditability must be treated as part of the deliverable if models are to support high-consequence operational decisions rather than remain laboratory demonstrations \cite{menke_geoenergy_ml_2026, abdellatif_benchmark_2026}. Significant technical challenges remain in upscaling. The scale gap between nano-CT (sample volumes of order 0.1~mm$^3$) and a mine (volumes of order $10^{10}$~m$^3$) spans roughly ten orders of magnitude, and the representative elementary volume, the minimum sample size at which measured physical properties are statistically stable, is poorly constrained for the texturally heterogeneous, compositionally variable ores most relevant to critical minerals. More broadly, the difficulty of moving from interface-scale dynamics to Darcy-scale description remains a fundamental problem across porous-media flow research \cite{berg_interface_darcy_2026}. ML upscaling methods have been validated mainly on structurally simpler systems such as sandstone reservoirs; their transferability to complex hard-rock ores, altered tailings, or heterogeneous battery black mass remains unproven at scale. In the near term, DRP is most likely to inform specific process decisions at the sample, particle, or core scale, rather than to produce mine-scale predictions from microstructural data alone. Credible site-scale relevance will require integration with core logging, geophysical surveys, and geostatistical frameworks, which is a cross-disciplinary investment extending well beyond imaging.

Institutional constraints are equally important. Synchrotron beamtime allocation is governed by scientific merit review rather than policy priority, and allocation cycles typically run to months rather than the weeks that mining and processing decisions often require. Industrial users face additional barriers: ore-characterisation data are commercially sensitive and companies are reluctant to place them in open repositories; sample preparation and transport to major facilities add cost and lead time; and the cost structures associated with proprietary beamtime access are difficult to absorb within the budgets of junior mining companies or specialist recyclers, which typically operate with limited R\&D resources. These conditions help explain why synchrotron-quality DRP has remained largely confined to academic and national laboratory settings. Any practical programme must therefore address access and economic incentives as directly as technical capability. Streamlined industrial access routes, block-booked beamtime for multi-user consortia, and cost-recovery models calibrated to SME participation are as important as the physical upgrades themselves. The tripartite consortium model proposed in Recommendation 1 will attract genuine industrial engagement only if IP arrangements, data-sharing obligations, and access economics are resolved at the call-design stage rather than deferred to project implementation.

\subsection{Infrastructure Priorities and Delivery Model}
Across exploration, processing, and circularity, the clearest infrastructure priority is investment in integrated geo-reactive platforms rather than another generation of standalone tomography instruments.

The delivery model should be distributed rather than centralised. Existing regional infrastructures, from NXCT in the UK to EXCITE and ECCSEL nodes across Europe, are the right place for screening, sample preparation, routine micro-CT, SEM-EDX automated mineralogy, training-data generation, and benchtop static or flow-through experiments \cite{manchester_nxct_2026, excite_access_2026, eccsel_strath_xct_2026}. Major facilities should focus on the measurements that genuinely require national-scale infrastructure: high-speed or high-energy tomography, correlative 3D chemistry, operando diffraction and spectroscopy, and in situ experiments that couple flow, temperature, pressure, and mechanics with GPU-based AI workflows and CPU/HPC reconstruction and simulation. Geological surveys and data infrastructures should host the reference datasets, standards, benchmark problems, and long-term curation, while industrial partners provide representative ores, tailings, brines, and recycling feedstocks together with process-validation routes.

Current US investment provides a useful benchmark. DOE is backing regional consortia, cross-laboratory platforms, hub programmes, and proving-ground style demonstration in parallel \cite{doe_regional_consortia_2025, doe_cmi_2025, doe_metallic_2024, doe_motf_launch_2025}. For DRP, the significance of this model lies in the way it links multimodal imaging and sensing to modelling, shared data environments, and pilot-scale decision-making within one translational framework. By contrast, UK and European investment is more visible in the form of excellent facilities, strategic-project designation, and project-by-project funding than in a fully articulated pathway linking regional sample handling, national multimodal experiments, shared data and compute, and pilot-scale validation \cite{eu_strategic_projects_2025, uk_vision_2035_2026}. The missing translation layer, linking multimodal imaging and sensing to modelling, shared data environments, and pilot-scale decision-making, is the central infrastructure gap that UK and European policy must address.

\begin{itemize}
\item \textbf{Correlative chemistry integrated with tomography:} routine hard X-ray workflows that register XRF-CT, XANES-CT, XRD-CT, and where feasible XES or Raman/IR measurements to the same 3D volume, so mineral texture and chemical state can be interpreted together rather than in separate campaigns.
\item \textbf{Reactive flow cells with synchronised effluent analysis:} experimental rigs for brines, acidic or alkaline lixiviants, and recycling leachates with inline measurements such as ICP-OES or ICP-MS, ion chromatography, pH, Eh, conductivity, and complementary optical spectroscopy, so that released species can be tied directly to evolving pore structure and flow pathways.
\item \textbf{Geo-relevant environmental control:} in situ platforms that combine flow and transport with temperature, confining pressure, differential stress, and mechanical loading, allowing experiments on fracture opening, permeability evolution, selective dissolution, and sorbent degradation under realistic operating conditions.
\item \textbf{AI-ready data and modelling infrastructure:} annotation workflows, benchmark datasets, versioned training and validation splits, and linked GPU and CPU/HPC compute environments so segmentation models, surrogates, and inverse models can be trained against reproducible experimental data rather than one-off campaigns.
\item \textbf{Translation infrastructure around the beamline:} reference ore suites, shared calibration standards, sample-preparation laboratories, interoperable data pipelines, and model-serving workflows so university microscopy centres can act as feeders into national facilities instead of isolated endpoints.
\end{itemize}

For critical minerals, the highest-value investment lies not in imaging alone, but in facility platforms that can observe structure, chemistry, and released products within a single time-coordinated workflow and feed those data directly into validated modelling environments. Routine co-registered 3D imaging, reacting-fluid analytics, outlet-chemistry measurements, GPU-based AI model updating, and CPU/HPC-scale reconstruction and simulation for geo-reactive systems remain unusual enough across major facilities to justify targeted investment. Such capability is needed to determine which minerals are liberated, dissolved, transported, precipitated, or lost under realistic processing conditions, and to carry those observations into predictive process models quickly enough to inform practice.

\begin{table}[htbp]
\centering
\footnotesize
\setlength{\tabcolsep}{6pt}
\renewcommand{\arraystretch}{1.25}
\rowcolors{2}{black!3}{white}
\begin{tabularx}{\textwidth}{L{0.18\textwidth}Y Y Y}
\toprule
\textbf{Existing Strength} & \textbf{Current Limitation} & \textbf{Added Investment} & \textbf{Expected Payoff} \\
\midrule
World-class XCT, nano-imaging, diffraction, and spectroscopy distributed across major facilities & Structure, chemistry, and fluid measurements are often split across separate campaigns or beamlines & Correlative endstations and cross-beamline workflows for XRF-CT, XANES-CT, XRD-CT, and operando imaging & Faster, decision-relevant links between ore texture, mineral chemistry, and process response \\
Strong in situ imaging capability for selected materials systems & Limited geo-specific sample environments for brines, lixiviants, tailings slurries, and black mass & Geo-reactive flow cells with corrosion-resistant fluidics and sample holders for ores, tailings, and recycling feedstocks & Direct evidence of dissolution, precipitation, permeability change, and selective recovery under realistic conditions \\
Advanced spectroscopy and diffraction with some gas-handling or operando capability & Weak routine coupling to synchronised outlet-fluid analytics for dissolved metals and reaction products & Inline ICP-OES/ICP-MS, ion chromatography, pH/Eh, conductivity, and optical probes connected to beamline experiments & Ability to link what leaves the sample chemically to what changes inside it structurally \\
Strong imaging output and growing digital capabilities, but fragmented metadata and annotation practice & Experimental datasets are rarely structured for reusable segmentation, surrogate modelling, or cross-site validation & AI-ready data standards, benchmark datasets, shared GPU and CPU/HPC compute environments, and model-evaluation workflows linked to facility experiments & Faster interpretation, more transferable models, and tighter coupling between experiments, simulation, and process decisions \\
Excellent national facilities but uneven local access & Regional centres can characterise samples, but often cannot feed data seamlessly into facility-scale experiments & Shared standards, calibration samples, sample-prep pipelines, interoperable data formats, and model-serving workflows from local labs to national repositories with access to national GPU and CPU/HPC resources & Lower transaction cost, more comparable datasets, broader participation from universities and SMEs, and faster translation from measurement to decision support \\
\bottomrule
\end{tabularx}
\caption{A practical investment model for critical-minerals DRP: build on existing strengths, close integration gaps, and emphasize platforms that connect structure, chemistry, and reactive transport.}
\label{tab:facility_gap_model}
\end{table}

Table~\ref{tab:facility_gap_model} indicates that the recommended investment is better conceived as a platform-integration programme than as a single instrument purchase. A phased approach is the most plausible. An initial phase should upgrade existing beamlines and laboratories with shared sample environments, outlet-chemistry analytics, reference materials, data engineering, and benchmark-ready modelling workflows running across GPU and CPU/HPC resources. A later phase should establish dedicated endstations only where demand, staff capacity, and demonstrator results justify them. The same logic applies to financing. As order-of-magnitude estimates, the initial phase would likely sit in a multi-million-pound or multi-million-euro upgrade category, while dedicated endstations with permanent staff, sample environments, GPU and CPU/HPC compute support, and data services would likely move into a low-tens-of-millions capital category.

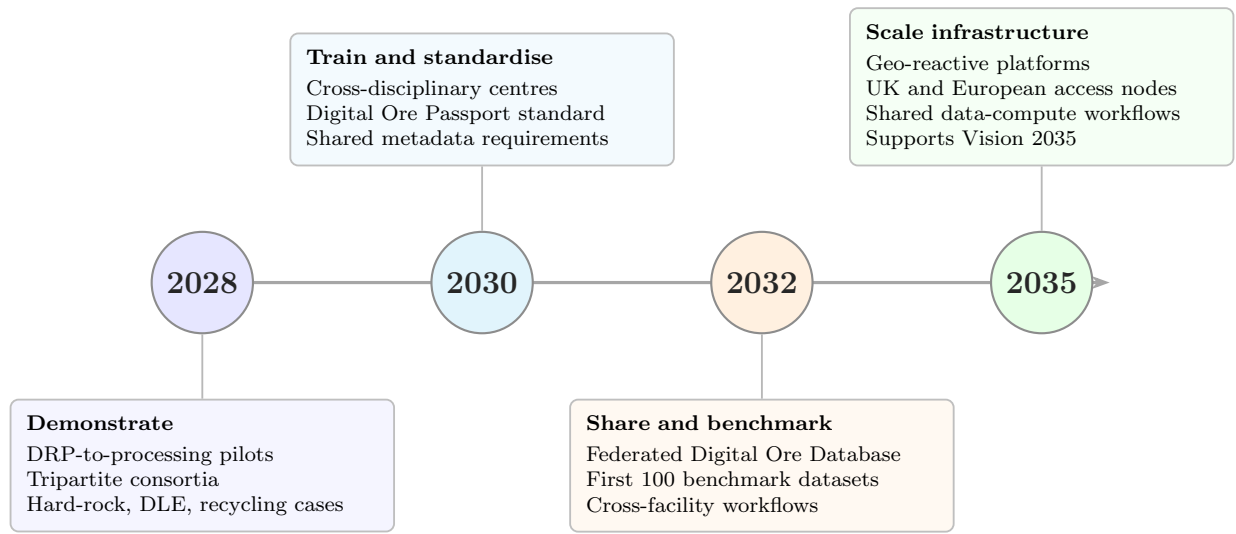
\begin{figure}[htbp]
\centering
\makebox[\textwidth][c]{%
\begin{tikzpicture}[
  year/.style={circle, draw=black!45, line width=0.8pt, minimum size=1.12cm,
               align=center, font=\bfseries\normalsize, text=black!85},
  card/.style={rectangle, rounded corners=3pt, draw=black!25, line width=0.7pt,
               minimum width=5.0cm, text width=4.65cm, align=left,
               font=\scriptsize, inner sep=6pt},
  line/.style={line width=1.2pt, draw=black!35},
  arrow/.style={-{Stealth[length=7pt,width=5pt]}, line width=1.2pt, draw=black!35},
  connector/.style={line width=0.7pt, draw=black!28}
]

\coordinate (y2028) at (2.45,0);
\coordinate (y2030) at (6.15,0);
\coordinate (y2032) at (9.85,0);
\coordinate (y2035) at (13.55,0);

\draw[line] (y2028) -- (y2032);
\draw[arrow] (y2032) -- (14.45,0);

\node[year, fill=blue!10] (n2028) at (y2028) {2028};
\node[year, fill=cyan!10] (n2030) at (y2030) {2030};
\node[year, fill=orange!12] (n2032) at (y2032) {2032};
\node[year, fill=green!10] (n2035) at (y2035) {2035};

\node[card, fill=blue!4, below=0.85cm of n2028] (c2028) {%
  \textbf{Demonstrate}\\[2pt]
  \mbox{DRP-to-processing pilots}\\
  \mbox{Tripartite consortia}\\
  \mbox{Hard-rock, DLE, recycling cases}
};

\node[card, fill=cyan!4, above=0.85cm of n2030] (c2030) {%
  \textbf{Train and standardise}\\[2pt]
  \mbox{Cross-disciplinary centres}\\
  \mbox{Digital Ore Passport standard}\\
  \mbox{Shared metadata requirements}
};

\node[card, fill=orange!5, below=0.85cm of n2032] (c2032) {%
  \textbf{Share and benchmark}\\[2pt]
  \mbox{Federated Digital Ore Database}\\
  \mbox{First 100 benchmark datasets}\\
  \mbox{Cross-facility workflows}
};

\node[card, fill=green!4, above=0.85cm of n2035] (c2035) {%
  \textbf{Scale infrastructure}\\[2pt]
  \mbox{Geo-reactive platforms}\\
  \mbox{UK and European access nodes}\\
  \mbox{Shared data-compute workflows}\\
  \mbox{Supports Vision 2035}\\
};

\draw[connector] (n2028.south) -- (c2028.north);
\draw[connector] (n2030.north) -- (c2030.south);
\draw[connector] (n2032.south) -- (c2032.north);
\draw[connector] (n2035.north) -- (c2035.south);

\end{tikzpicture}%
}
\caption{Proposed implementation roadmap for a UK-European critical-minerals DRP programme. Near-term demonstrators provide the evidence base for cross-disciplinary training and a Digital Ore Passport standard, followed by a federated Digital Ore Database, shared benchmark datasets, integrated geo-reactive infrastructure, and mature decision-support workflows aligned with the UK's 2035 domestic-production ambition.}
\label{fig:implementation_roadmap}
\end{figure}

Figure~\ref{fig:implementation_roadmap} translates these infrastructure priorities into a staged delivery roadmap aligned with the near-term recommendations, the EU's 2030 CRMA benchmarks, and the UK's 2035 domestic-production ambition.

\FloatBarrier
\clearpage
\newgeometry{a4paper, top=0.55in, bottom=0.55in, left=0.68in, right=0.68in}

The policy implication is that DRP should be governed and funded as pre-competitive implementation infrastructure. That means designing public funding calls, data standards, access rules, and facility investments so that evidence from one ore, brine, tailings deposit, or recycling stream can inform wider policy and industrial decision-making. The following recommendations identify actions that could be initiated within a single public funding cycle:

\begin{tcolorbox}[enhanced, colback=gray!3!white, colframe=blue!40!black,
title=\textbf{Key Policy Recommendations}, colbacktitle=blue!40!black,
drop fuzzy shadow, fonttitle=\bfseries\normalsize, arc=1.2mm, boxrule=0.7pt,
boxsep=0.6mm, left=0.9mm, right=0.9mm, top=0.7mm, bottom=0.7mm,
toptitle=0.7mm, bottomtitle=0.7mm, fontupper=\normalsize,
before upper={\setlength{\parskip}{0.8pt}\setlength{\parindent}{0pt}\setlength{\baselineskip}{11.8pt}}]
\textbf{1. Launch Translational Demonstrators by 2028:}\\
UKRI, Horizon Europe, and national innovation agencies should create DRP-to-processing pilot calls with tripartite consortia spanning imaging science, an orebody owner or recycler, and process scale-up. A first portfolio of three to five projects should include one hard-rock liberation case, one brine or DLE case, and one recycling or tailings case. Each should move from characterisation to pilot-plant decision within 24--36 months and deliver a public benchmark dataset, a documented recovery or processing decision, and a comparison with conventional characterisation.

\vspace{2.4mm}
\textbf{2. Build Workforce Capacity with Cross-Disciplinary Training Centres:}\\
The UK and Europe should fund doctoral and industrial training centres at the intersection of imaging, machine learning, mineral processing, use, and recycling. These centres should be tied to demonstrator projects and train people able to move between beamlines, laboratories, software teams, geological surveys, and pilot plants. Two to three centres, collectively supporting at least 40 jointly supervised researchers by 2030, would form a meaningful initial cohort.

\vspace{2.4mm}
\textbf{3. Define a ``Digital Ore Passport'' Standard by 2030:}\\
The European Critical Raw Materials Board, relevant UK funding and geological-survey institutions, EuroGeoSurveys, BGS, national geological surveys, and major imaging facilities should publish a common minimum data specification for publicly supported mining, recycling, and tailings projects \cite{eu_crm_board_2026, bgs_uk_crm_2022, uk_vision_2035_2026}. It should include multi-modal imaging metadata, mineralogical ground truth, uncertainty estimates, machine-readable process descriptors, and standardized influent and effluent chemistry. Adoption should be required for publicly funded pilot projects from 2030 onward.

\vspace{2.4mm}
\textbf{4. Stand Up a Federated UK-European Digital Ore Database by 2032:}\\
The European Commission, UK agencies, national geological surveys, and major imaging facilities should create a shared repository with open metadata, benchmark datasets, and APIs for model development. The practical route is to build through EGDI, ensure interoperability with RMIS, and add a linked UK contribution rather than isolated parallel platforms \cite{egdi_2026, rmis_2026, bgs_uk_crm_2022}. Deposition should be a public-funding condition, long-term curation should be resourced, and the schema should serve process engineers, geometallurgists, and ML developers. A first milestone should be 100 benchmark datasets, shared reference materials, and at least three validated cross-facility workflows.

\vspace{2.4mm}
\textbf{5. Build Integrated Geo-Reactive Endstations by 2035:}\\
Major facilities in the UK and Europe should co-invest in shared endstations combining hard X-ray imaging, chemically sensitive mapping, synchronised outlet-fluid analysis, and GPU/CPU/HPC workflows. Platforms should support brines, acidic and alkaline lixiviants, elevated temperatures and pressures, controlled mechanical loading, and inline chemistry. The first stage should upgrade existing beamlines and laboratories; dedicated endstations should follow where demand and demonstrator outcomes justify them. A practical target is one UK platform and one to two continental European platforms, operated through coordinated access and common data standards.
\end{tcolorbox}
\clearpage
\restoregeometry

\section{Conclusion}
Critical-minerals policy is entering an implementation phase. The central question is no longer only how to set targets for domestic extraction, processing, and recycling, but how to generate the evidence needed to determine which resources and secondary feedstocks can be developed efficiently, responsibly, and at acceptable environmental cost. Many of the most consequential decisions in critical-minerals development, from orebody evaluation to comminution, leaching, and recycling, depend on three-dimensional structural and reactive information that bulk assays and section-based methods cannot fully resolve. DRP can supply that evidence by combining multiscale imaging, correlative chemistry, physics-based simulation, AI-assisted interpretation, and access to GPU and CPU/HPC compute to characterise both ores and engineered porous materials. In UK and European settings, its value lies not only in stronger supply security and process efficiency, but also in making lower-impact development pathways easier to identify and evaluate. The highest-value facility investment is therefore not another standalone imaging instrument, but integrated geo-reactive platforms, shared data standards, and demonstrator programmes that connect measurement to policy-relevant decisions. The first wave should centre on one brine or DLE demonstrator, one hard-rock liberation demonstrator, and one recycling or tailings demonstrator, all linked from the outset to common data standards and a federated UK-European digital ore infrastructure.

\section*{Acknowledgements}
The views expressed in this paper are those of the authors and do not necessarily reflect the views or policy positions of the International Energy Agency or its member countries.

\section*{Author Contributions}
H.P.M.: Conceptualization, Writing -- original draft. A.S.: Investigation, Writing -- review \& editing. M.R.: Investigation, Writing -- review \& editing.

\end{document}